\documentclass[preprint,10pt,3p,authoryear]{elsarticle} % other options include
\usepackage{dcolumn}
\usepackage{bm}
\usepackage{graphicx}
\usepackage[dvips]{color}
\usepackage{epsfig}
\usepackage{float}
\usepackage{subfig}
\usepackage{verbatim}
\usepackage{dcolumn}% Align table columns on decimal point
\usepackage{setspace}
\usepackage{lineno}
\usepackage{natbib}
\usepackage{rotating} %Need this for sidewaystable
\newcommand{\degree}{\ensuremath{^\circ}}

\biboptions{compress}

\begin{document}

\title{Measurement of airborne fission products in Chapel Hill, NC, USA from the Fukushima Dai-ichi reactor accident}

%ELSARTICLE Author Style
\author[unc,tunl]{S.~MacMullin\corref{cor}}
\ead{spm@physics.unc.edu}
\author[unc,tunl]{G.K.~Giovanetti}
\author[unc,tunl]{M.P.~Green}
\author[unc,tunl]{R.~Henning}
\author[unc]{R.~Holmes\fnref{fn1}}
\author[unc,tunl]{K.~Vorren}
\author[unc,tunl,ornl]{J.F.~Wilkerson}

\cortext[cor]{Correspoding author}
\fntext[fn1]{Current address: Department of Physics, University of Illinois, Urbana, IL, 61801 USA}
\address[unc]{Department of Physics and Astronomy, University of North Carolina, Chapel Hill, NC , 27599 USA}
\address[tunl]{Triangle Universities Nuclear Laboratory, Durham, NC, 27708 USA}
\address[ornl]{Oak Ridge National Laboratory, Oak Ridge, TN, 37830 USA}

\begin{keyword}
    Radioactive fallout, Fukushima nuclear accident, Fission product radionuclides, Gamma spectroscopy
\end{keyword}

\begin{abstract}
We present measurements of airborne fission products in Chapel Hill, NC, USA, from 62 d following the March 11, 2011, accident at the Fukushima Dai-ichi nuclear power plant. Airborne particle samples were collected daily in air filters and radio-assayed with two high-purity germanium (HPGe) detectors. The fission products \textsuperscript{131}I and \textsuperscript{137}Cs were measured with maximum activity concentrations of 4.2 $\pm$ 0.6 mBq/m\textsuperscript{3} and 0.42 $\pm$ 0.07 mBq/m\textsuperscript{3} respectively. Additional activity from \textsuperscript{131,132}I, \textsuperscript{134,136,137}Cs and \textsuperscript{132}Te were measured in the same air filters using a low-background HPGe detector at the Kimballton Underground Research Facility (KURF).
\end{abstract}
\maketitle

%_______________________________________________ACTUAL TEXT___________________________________________
\section{Introduction}

On March 11, 2011, reactors at the Fukushima Dai-ichi nuclear power plant in Japan (37\degree45' N, 141\degree27' E) were shut down following a 9.0 magnitude earthquake. Emergency diesel generators were activated to power water pumps needed to cool the reactors and prevent intensely radioactive nuclear fuel from overheating and damaging the reactor containment vessels. Shortly after the earthquake, the plant was struck by a 14-m tsunami which flooded the electrical building, disabling the emergency generators~\citep{wnn}. Fires, explosions and possible partial core meltdowns released radioactive fission products into the atmosphere. On April 20, 2011, the International Atomic Energy Agency (IAEA) reported \textsuperscript{131}I deposition in Japan ranging from 1.8 to 368 Bq/m\textsuperscript{2} in 13 prefectures and \textsuperscript{137}Cs deposition ranging from 2.4 to 160 Bq/m\textsuperscript{2} in seven prefectures~\citep{iaea}.

Uncertainty surrounded the development of the situation at Fukushima as plant staff struggled to restore power and to adequately cool the reactors. A limited amount of information and measurements relevant to the release of radioactive material had been available. Considering global interest in possible public health effects and local impact on our physics program that relies on ultra-low radioactive background detectors, we began collecting airborne particle samples and monitoring for fission products on March 17, 2011. 

Airborne fission products, including \textsuperscript{131,132}I, \textsuperscript{134,137}Cs, \textsuperscript{132}Te and \textsuperscript{133}Xe were first detected on the west coast of the United States on March 16, 2011, and reached a maximum activity concentration of 4.4 $\pm$ 1.3 mBq/m\textsuperscript{3} of \textsuperscript{131}I 3 d after the first detection of fission products~\citep{Bow11,Leo11}. Radioactivity has also been measured in Europe~\citep{Mas11}, western Japan~\citep{Fus11}, Russia~\citep{Bol11}, in rainwater in the United States~\citep{Nor11} and is continuously monitored in France~\citep{IRSN11}.

We first detected airborne fission products in Chapel Hill between 20:00 UTC on March 18, 2011, and 20:00 UTC on March 19, 2011. We measured a maximum activity concentration of 4.2 $\pm$ 0.6 mBq/m\textsuperscript{3} of \textsuperscript{131}I in the interval between March 29, 2011 and March 30, 2011. The time dependence of \textsuperscript{131}I and \textsuperscript{137}Cs activity seems to have been dominated by local rain, but we are able to draw some conclusions about the nature of the reactor accident and the transport of fission products around the world.

\section{Materials and Methods}

The air sampling pump and filter were housed on the roof of Phillips Hall at the University of North Carolina at Chapel Hill (35\degree55'N, 79\degree2' W, 150 m elevation). The pump is a Staplex model TFIA 110-125 V DC/AC with built-in flow meter and an 8'' $\times$ 10'' filter holder assembly \citep{samplers}. With a filter in place, the flow rate of the air sampler was measured to be 2050 m\textsuperscript{3}/d. Air filters collected airborne particles for approximately 24 h, then were removed, folded into 5 cm $\times$ 10 cm rectangles and sealed in nylon bags. A total of 38 air filters were collected over 62 d following the March 11, 2011 earthquake.

Two 5\% relative efficiency to NaI (R.E.) high-purity germanium (HPGe) detectors were used to identify characteristic gamma rays from radioactive fission products. The detectors were setup horizontally with a 10 mm separation between their end caps. Bagged filters were placed between the end caps of the detectors for assay. The detectors were surrounded by a lead shield of 5-10 cm thickness to reduce backgrounds from cosmic rays and natural radiation. 

Ten filters which had already been measured using the setup described above were assayed at the Kimballton Underground Research Facility (KURF) in order to search for more exotic, long-lived fission products. KURF is located in the Kimballton mine near Ripplemead, VA, which provides shielding equivalent to 1450 m of water from cosmic-ray induced backgrounds. The detectors in this counting facility have a background rate about 40 times less than the above-ground detectors in the region of 40 -- 2700 keV. All 10 filters were tightly fit into a Marinelli beaker and were assayed simultaneously in the `VT-1' detector which is a 35\% R.E. coaxial HPGe detector with very low background (ORTEC LLB). For technical specifications on the counting systems and facility see~\citet{Fin10}.

A single filter was also measured using the VT-1 detector to determine the gamma-ray detection efficiency of the surface based counting system. The efficiency of the VT-1 detector had previously been determined using an existing Monte Carlo simulation that had been calibrated using point sources of known activity spanning wide gamma-ray energy range in multiple locations surrounding the detector. The detection efficiencies of the above-ground detectors were determined by comparing \textsuperscript{131}I and \textsuperscript{137}Cs count rates to VT-1 measurements of the same filter. The relative efficiencies were also verified by comparing the efficiency of a \textsuperscript{133}Ba (E$_\gamma$ = 356 keV) point source on the top of the cryostat of the VT-1 detector to a similar source of known activity in the same position on the cryostats of the above-ground detectors.

To determine the filter efficiency for air particles containing \textsuperscript{131}I and \textsuperscript{137}Cs, two filters were placed in the air sampler in series and the activities of the two filters were compared. Using this method, the filter efficiency for particles carrying \textsuperscript{131}I was determined to be 87\% $\pm$ 5\%. The filter efficiency for air particles containing \textsuperscript{137}Cs was determined to be 98.7\% $\pm$ 0.6\%. The manufacturer quotes a filter efficiency of 99.98\% for particles down to 0.3 $\mu$m \citep{filters}. It should be noted that the measured \textsuperscript{131}I only represents particulate species collected in the air filters. This accounts for only about 50\% of the total \textsuperscript{131}I in the air. The rest is distributed in gases such as I$_2$ and methyl iodide \citep{Per90}.

%TFAGF810 8'' $\times$ 10'' glass fiber filters from Staplex

The particulate activity concentration from a given isotope can be calculated as:

\begin{equation}
A_{air} = \frac{R}{\epsilon_{det}\epsilon_{filter}}\frac{1}{Qt_{samp}}
\end{equation}

\noindent where $\epsilon_{det}$ is the absolute gamma-ray detection efficiency, $\epsilon_{filter}$ is the air filter efficiency, $Q$ is the flow rate of the air sampler, t$_{samp}$ is the time duration of air sampling and $R$ is the count rate of the filter at the end of the sampling period. $R$ has been corrected for any delay between sampling and counting using the known lifetimes of the relevant isotopes. We did not correct for decays during the sampling period.

\section{Results and Discussion}

\begin{figure*} [htp]
	\centering
	\subfloat[]{\epsfig{file=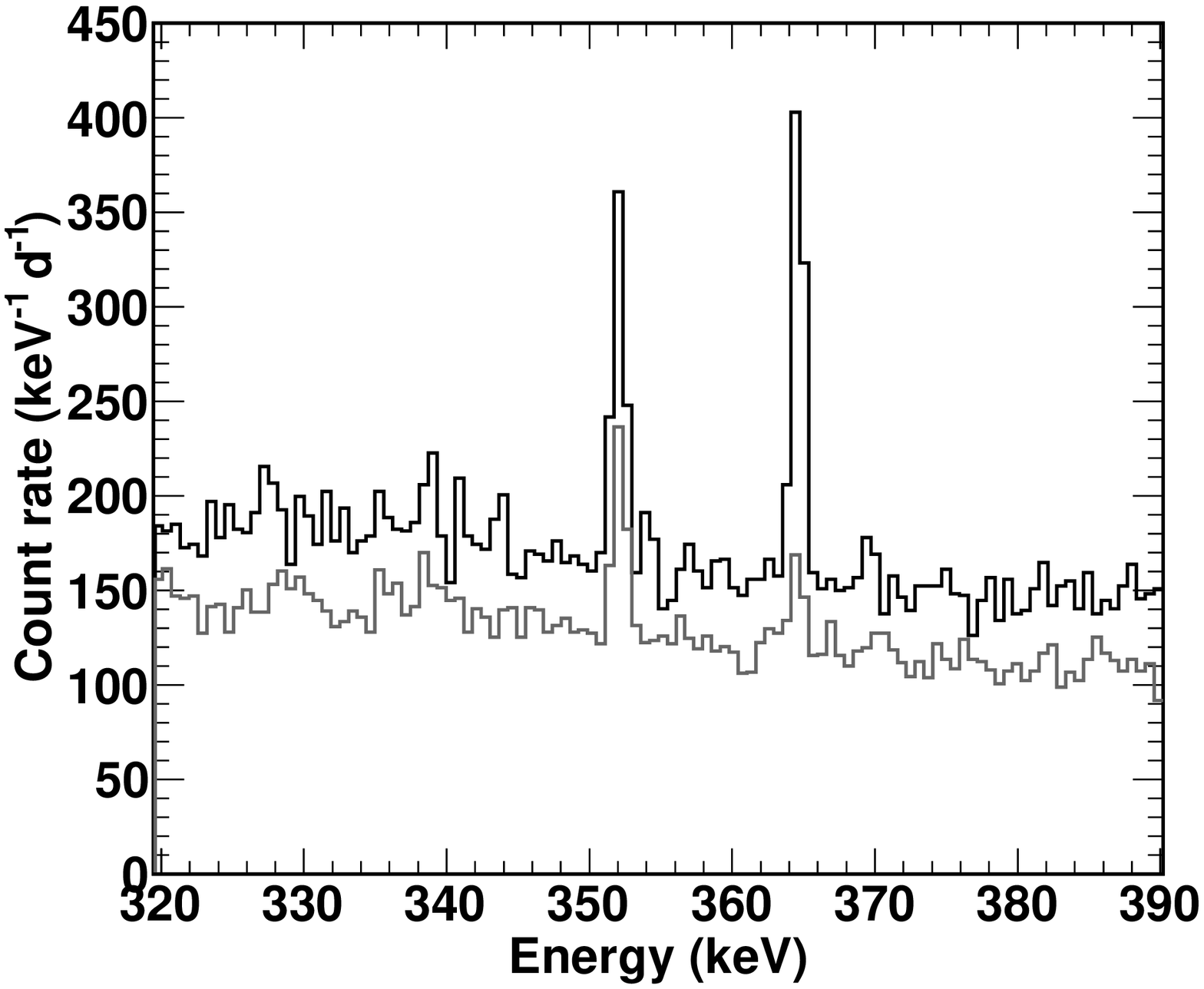, width=.45\textwidth}}
	\subfloat[]{\epsfig{file=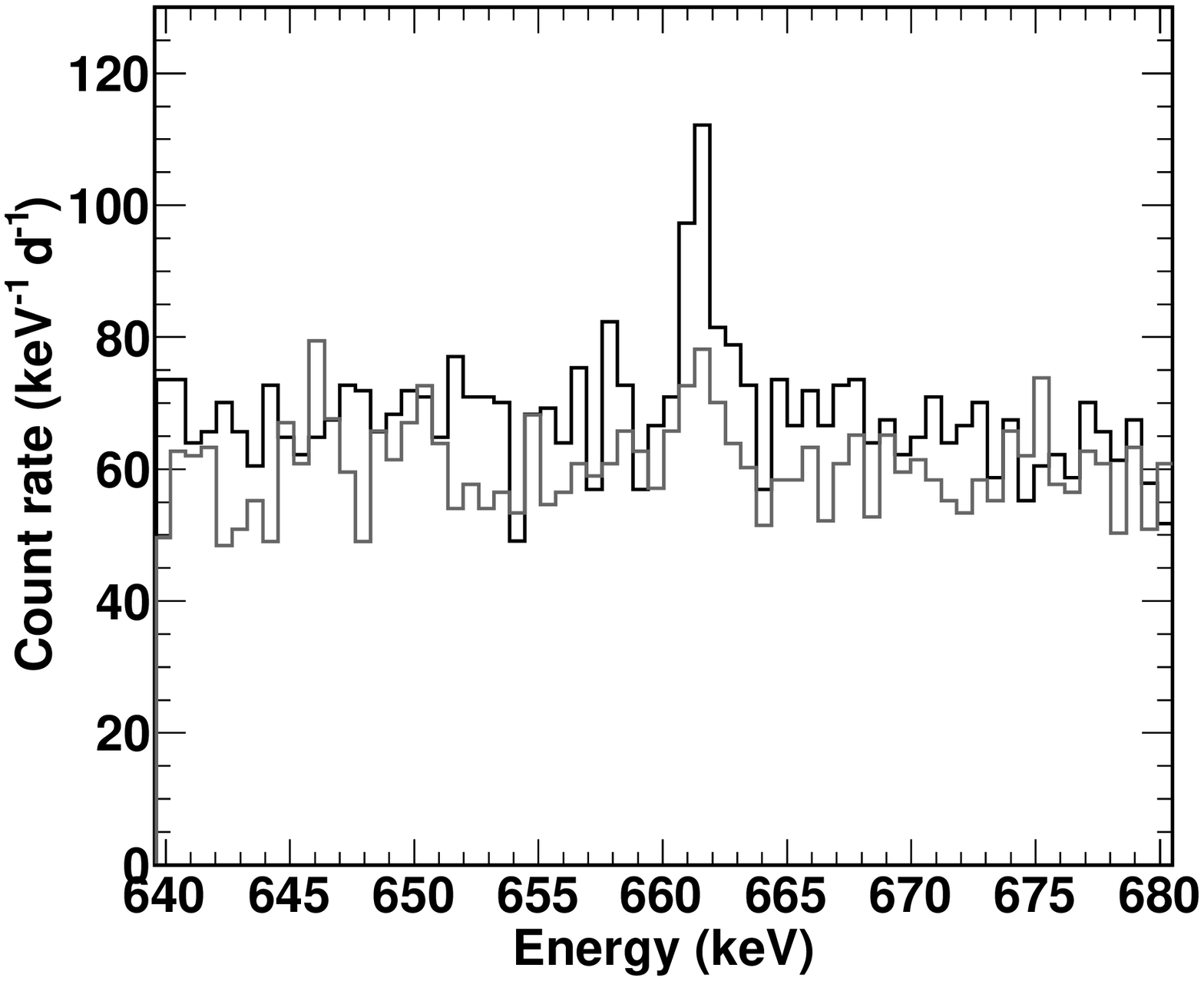, width=.45\textwidth}}\\
	\caption{Comparison of gamma-ray spectra from the filter sampled from March 18 - 19, 2011 (gray) and March 20 - 21, 2011 (black) showing an increase in fission fragments. (a) The dominant fission fragment peak is at 364 keV from the decay of \textsuperscript{131}I. The peak at 352 keV is from \textsuperscript{214}Pb, which comes from \textsuperscript{222}Rn in the air. (b) Fission fragment peak from \textsuperscript{137}Cs at 662 keV.}
	\label{fig:arrival}
\end{figure*}

\begin{figure*}
\centering
\includegraphics[width=.9\textwidth, angle=0]{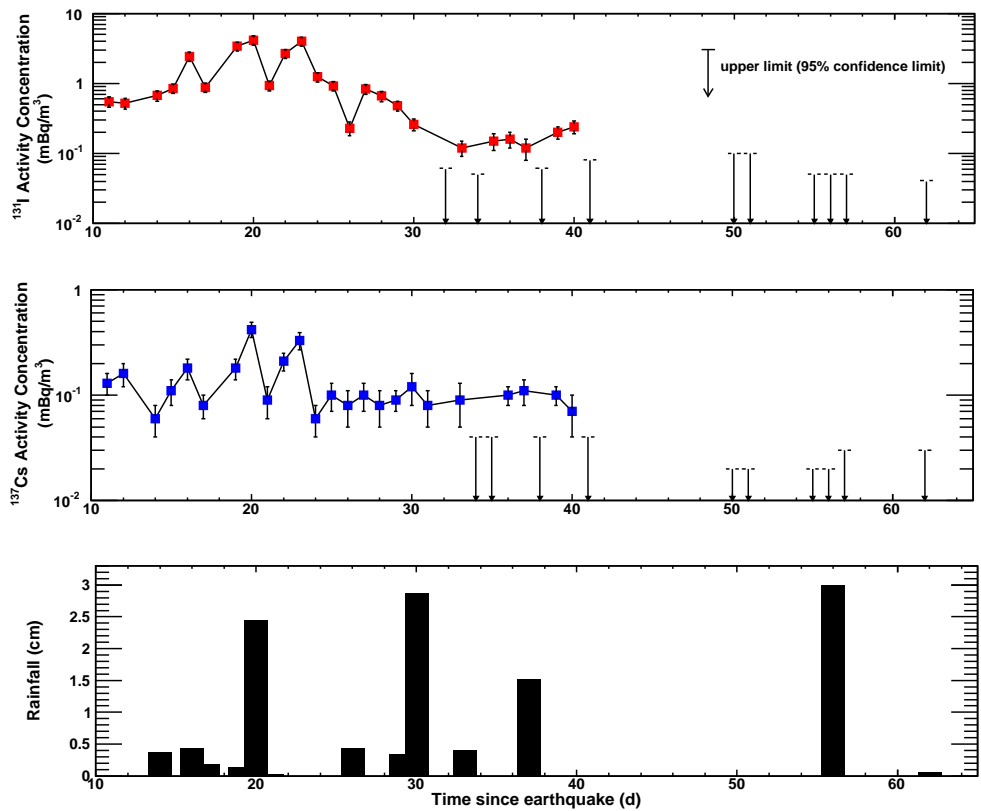}
\caption{The time dependence of \textsuperscript{131}I and \textsuperscript{137}Cs activity and local rainfall. Time is shown in days after the March 11 earthquake (rounded to the nearest day). Each activity measurement corresponds to the day the filter was removed from the air sampler. Days when the fission product activity was below the detection sensitivity are reported as upper limits (95\% confidence limit). The data may also be found in Table \ref{samplingdata}.}
\label{time}
\end{figure*}

The first filter was placed in the air sampler on March 17, 2011, and collected particles for 24 h. Although there are many nuclear facilities in the United States including the Shearon Harris nuclear power plant about 30 km from the sampling site~\citep{nrc2}, no fission products were detected in this filter and all visible gamma-ray peaks could be attributed to known background sources. Filters from subsequent days produced clear peaks attributable to \textsuperscript{131}I and \textsuperscript{137}Cs as seen in Fig.~\ref{fig:arrival}. The sampling dates and measured air activity concentrations for the first 62 d of data are listed in Table \ref{samplingdata}. A plot of the \textsuperscript{131}I and \textsuperscript{137}Cs activity concentrations measured over the first 40 d is shown in Fig.~\ref{time}.

\begin{table*}
\small
  \caption{\ Measured \textsuperscript{131}I and \textsuperscript{137}Cs air activity concentrations during the first 62 d of air sampling. Filters collected air particles for approximately 24 h.}
  \label{samplingdata}
  \begin{tabular*}{.99\textwidth}{@{\extracolsep{\fill}}lllll}
    \hline
    \\
Filter sampling                 & Days since  & Precipitation\textsuperscript{b} &  \textsuperscript{131}I  activity & \textsuperscript{137}Cs activity\\
  end date\textsuperscript{a} & earthquake  &       (cm)                            &  concentration    & concentration  \\
		                        & 			&					             &	(mBq/m\textsuperscript{3})  &	(mBq/m\textsuperscript{3}) 		\\
    \hline
March 19 	 &9				& 0.00   					  & 0.36 $\pm$ 0.07 		& 0.11 $\pm$ 0.03\\
March 21   &11			& 0.00    					  & 0.55 $\pm$ 0.09 		& 0.13 $\pm$ 0.03\\
March 22 	 &12			& 0.00   					  & 0.52 $\pm$ 0.09 		& 0.16 $\pm$ 0.04\\
March 24   &14			& 0.38 					  & 0.67 $\pm$ 0.11 		& 0.06 $\pm$ 0.02\\
March 25 	 &15			& 0.00    					  & 0.85 $\pm$ 0.13 		& 0.11 $\pm$ 0.03\\
March 26 	 &16			& 0.44 					  & 2.42 $\pm$ 0.37 		& 0.18 $\pm$ 0.04\\
March 27 	 &17			& 0.18 					  & 0.88 $\pm$ 0.14 		& 0.08 $\pm$ 0.02\\
March 29 	 &19			& 0.13    				  & 3.40 $\pm$ 0.51 		& 0.18 $\pm$ 0.04\\
March 30 	 &20			& 2.44 					  & 4.15 $\pm$ 0.62 		& 0.42 $\pm$ 0.07\\
March 31 	 &21			& 0.03 					  & 0.93 $\pm$ 0.15 		& 0.09 $\pm$ 0.03\\
April 1 	 &22			& 0.00     				  & 2.66 $\pm$ 0.40 		& 0.21 $\pm$ 0.04\\
April 2 	 &23			& 0.00    					  & 3.99 $\pm$ 0.60 		& 0.33 $\pm$ 0.06\\
April 3 	 &24			& 0.00     				  & 1.24 $\pm$ 0.19 		& 0.06 $\pm$ 0.02\\
April 4 	 &25			& 0.00    					  & 0.92 $\pm$ 0.14 		& 0.09 $\pm$ 0.03\\
April 5 	 &26 			& 0.43   				  & 0.23 $\pm$ 0.05 		& 0.08 $\pm$ 0.02\\
April 6 	 &27			& 0.00     				  & 0.83 $\pm$ 0.13 		& 0.10 $\pm$ 0.03\\
April 7 	 &28			& 0.00 	 				  & 0.66 $\pm$ 0.11 		& 0.08 $\pm$ 0.03\\
April 8    &29			& 0.33  				  & 0.48 $\pm$ 0.08 		& 0.09 $\pm$ 0.02\\
April 9 	 &30			& 2.87     				  & 0.26 $\pm$ 0.05 		& 0.12$\pm$ 0.04\\
April 10 	 &31			& 0.00    					  & $<$ 0.06 		& 0.08 $\pm$ 0.03\\
April 11 	 &32			& 0.00   			 		  & $<$ 0.06 		& $<$ 0.04\\
April 12 	 &33			& 0.41    				  & 0.12 $\pm$ 0.03 		& 0.09 $\pm$ 0.04\\
April 13   &34			& 0.00    					  & $<$ 0.05 		& 		$<$ 0.04\\
April 14   &35			& 0.00    					  & 0.15 $\pm$ 0.04 		& $<$ 0.04\\
April 15   &36			& 0.00    					  & 0.16 $\pm$ 0.04 		& 0.10 $\pm$ 0.02\\
April 16   &37			& 1.52 					  & 0.12 $\pm$ 0.04 		& 0.11 $\pm$ 0.03\\
April 17   &38		    & 0.00   					  & $<$ 0.06         		& $<$ 0.04\\
April 18   &39			& 0.00    					  & 0.20 $\pm$ 0.04 		& 0.10 $\pm$ 0.02\\
April 19   &40			& 0.00    					  & 0.24 $\pm$ 0.05 		& 0.07 $\pm$ 0.03\\
April 20   &41			& 0.00    					  & $<$ 0.08 		& $<$ 0.04\\
April 29  & 50			& 0.00							& $<$ 0.1					& $<$ 0.02 \\
May 1    & 51			& 0.00							& $<$ 0.1					& $<$ 0.02 \\
May 4  	 &55			& 0.00    					  & $<$ 0.05         		& $<$ 0.02\\
May 5    &56			& 3.00 					  & $<$ 0.05        		& $<$ 0.02\\
May 6    &57			& 0.00   					  & $<$ 0.05         		& $<$ 0.03\\
May 11 	 &62			& 0.05 					  & $<$ 0.04        		& $<$ 0.03\\
    \hline
  \end{tabular*}
Uncertainties from counting statistics are $\pm$ 1 standard deviation/error. Upper limits are reported at 95\% confidence limit. \\
\textsuperscript{a}The air filters which ended sampling on March 18, March 20 and March 23 were assayed in a single detector counting system and are not included here. The March 18 filter had no detectable fission product isotopes. Filters were sampled less frequently and counted for longer durations as the activity approached our detection limit.\\
\textsuperscript{b}\citep{wund}
\end{table*}

\begin{figure*} [htp]
	\centering
	\subfloat[]{\epsfig{file=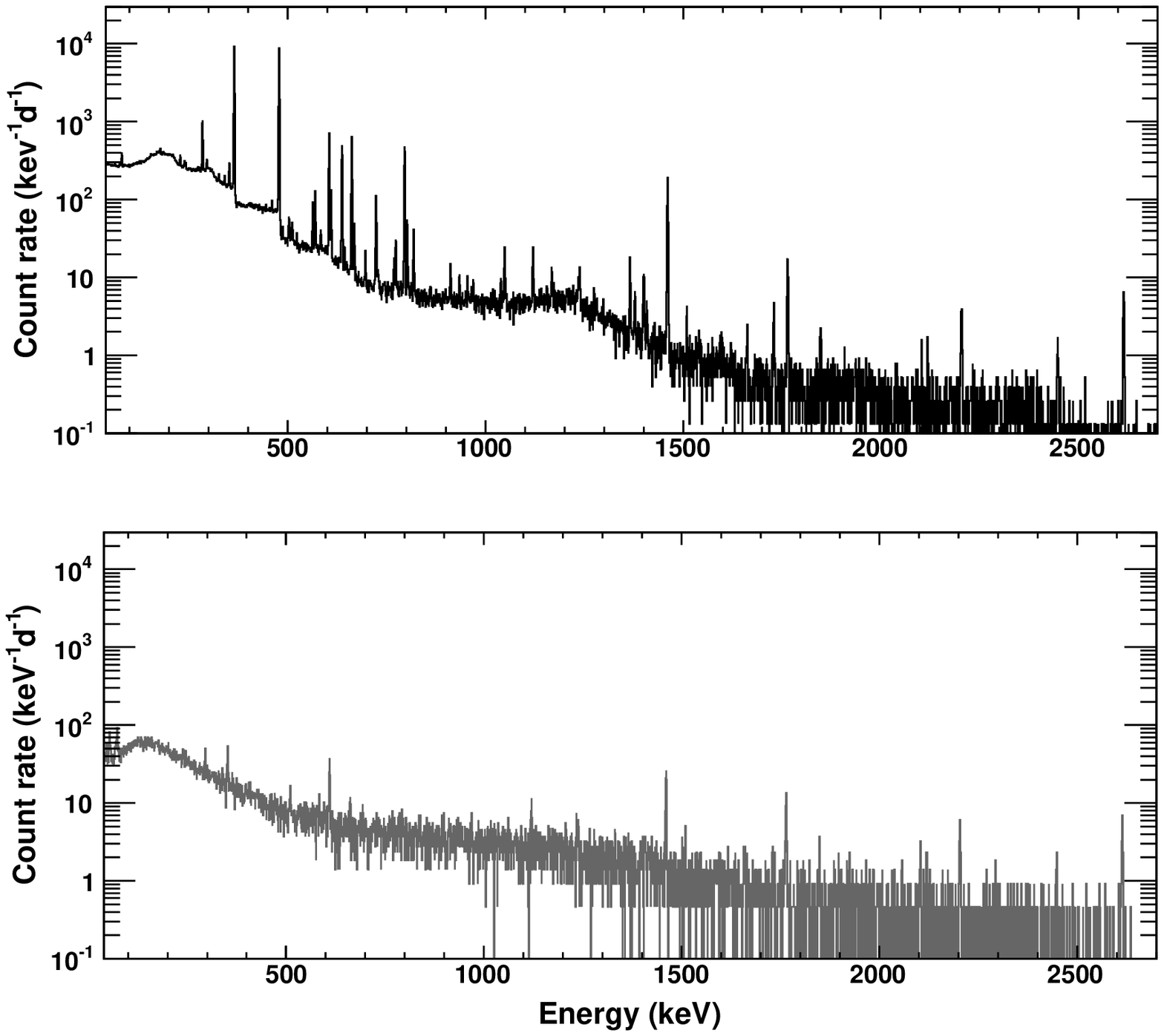, width=.90\textwidth}}\\
	\subfloat[]{\epsfig{file=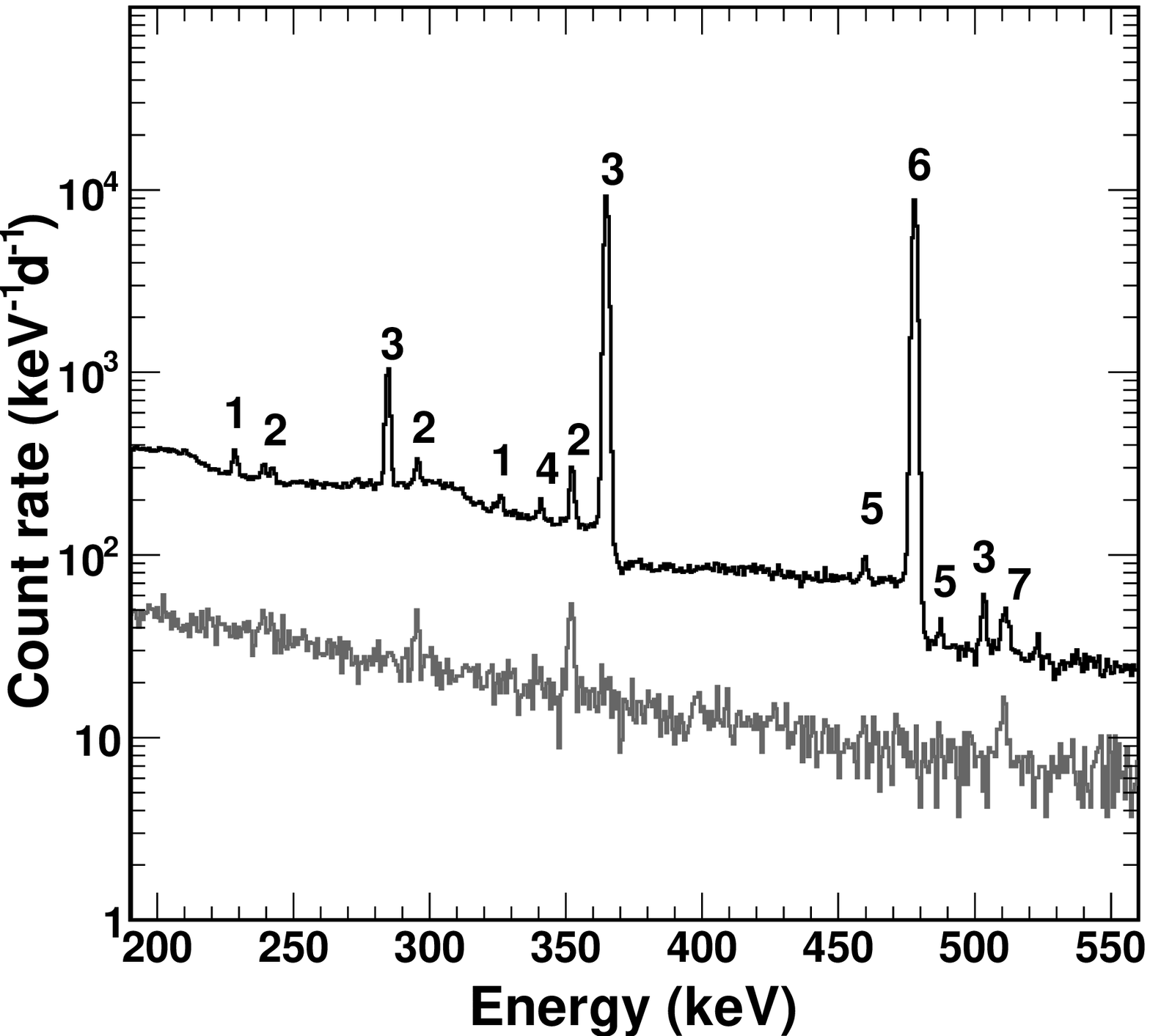, width=.45\textwidth}}
	\subfloat[]{\epsfig{file=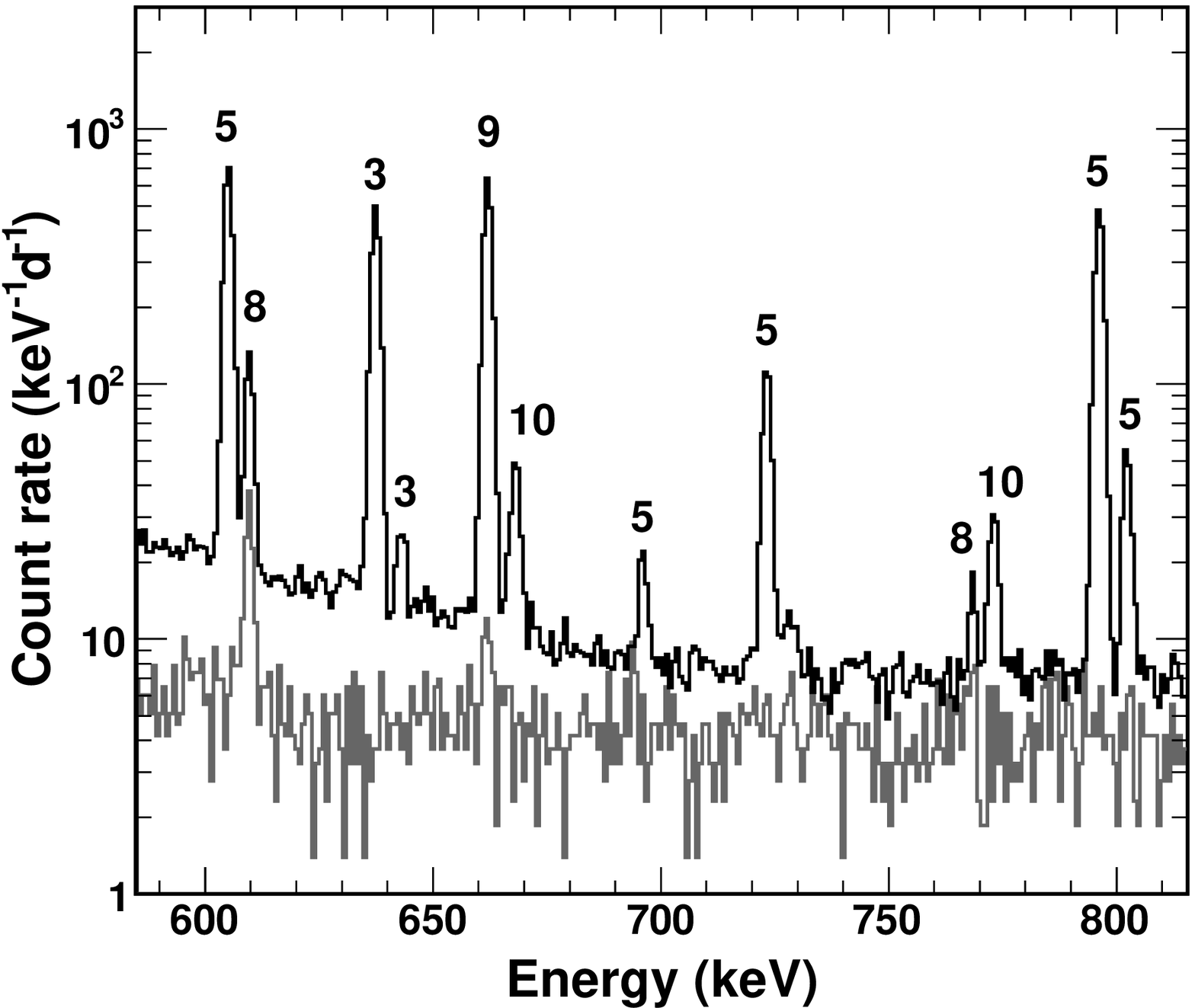, width=.45\textwidth}}\\
	\caption{(a) Gamma-ray spectrum for the 10 filters counted simultaneously (black) and background spectrum (gray) in VT-1. (b),(c) Two regions of the VT-1 filter and background spectra where various fission fragment and background peaks are visible. Number labels correspond to the isotopes listed in Table~\ref{tab:kurf}.}
	\label{fig:kurf}
\end{figure*}

Additional fission product isotopes were detected by simultaneously counting filters from the first 10 d of air sampling  at KURF. Fig.~\ref{fig:kurf} shows gamma-ray spectra from this measurement where peaks from the isotopes \textsuperscript{132}Te and \textsuperscript{136}Cs, which could not be seen in the above-ground detectors, are visible. The fission products and gamma-ray energies are listed in Table~\ref{tab:kurf}. Two individual filters were also counted at KURF to measure the activity of fission products that were below the detection limit of the above-ground detectors. The activity concentrations measured in these filters are listed in Table~\ref{tab:kurfresults}.

\begin{table*}
\small
  \caption{Fission products and backgrounds detected at KURF for peaks seen in the part of the spectra displayed in Fig.~\ref{fig:kurf}.}
  \label{tab:kurf}
  \begin{tabular*}{.99\textwidth}{@{\extracolsep{\fill}}lllll}
    \hline
    \\
Isotope &  Fig.~\ref{fig:kurf} Label  &Source & $T_{1/2}$ & Gamma-rays detected (keV)   \\
\hline
\\
\textsuperscript{132}Te & 1                      & fission product  & 76.9 h     & 228, 326\\
\textsuperscript{214}Pb & 2                      & background from \textsuperscript{222}Rn        & 26.8 min  & 242, 295. 352\\
\textsuperscript{131}I  & 3                       & fission product  & 8.03 d      & 80.2, 176, 284, 364, 503, 637, \\
		&			&				&			& 643, 723 \\
\textsuperscript{136}Cs & 4                      & fission product  & 13.0 d     & 340, 818, 1048, 1235\\
\textsuperscript{134}Cs & 5                      & fission product  & 2.07 y     & 458, 487, 563, 569, 605, 695, \\
		&			&				&			& 796, 802, 1039, 1168, 1365 \\
\textsuperscript{7}Be   & 6                      & cosmogenic       & 53.2 d      & 477\\
\textsuperscript{208}Tl & 7                      & background from \textsuperscript{232}Th        & 3.05 min  & 583, 2614\\
\textsuperscript{214}Bi & 8                      & background from \textsuperscript{222}Rn        & 19.9 min  & 609, 768, 1120, 1765, 2204\\
\textsuperscript{137}Cs & 9                      & fission product  & 30.1 y     & 661\\
\textsuperscript{132}I & 10                      & fission product  & 137.7 min  & 667, 773, 954\\
    \hline
  \end{tabular*}
\end{table*}

\begin{sidewaystable*}
\small
  \caption{\ Assay results from filters measured at KURF. The filter efficiency for \textsuperscript{132}Te is assumed to be 99.98\%. All activity concentrations are in units of mBq/m\textsuperscript{3}.}
  \label{tab:kurfresults}
\begin{tabular*}{.99\textwidth}{@{\extracolsep{\fill}}lllllll}
 \hline
\\
Filter Sample & \textsuperscript{132}Te 	 & \textsuperscript{131}I 		& \textsuperscript{132}I 		   		& \textsuperscript{134}Cs 		& \textsuperscript{136}Cs 		& \textsuperscript{137}Cs\\
End Date			   & 				 &					&					  		&			  		&					&   		\\
\hline
\\
March 20			   & 0.08 $\pm$ 0.03 & 0.57 $\pm$ 0.06  & 0.024 $\pm$ 0.004   		& 0.038 $\pm$ 0.004 & 0.007 $\pm$ 0.002 & 0.043 $\pm$ 0.005 \\
March 29			   & 0.05 $\pm$ 0.01 &  4.2 $\pm$ 0.5   & 0.08 $\pm$ 0.02    		& 0.020 $\pm$ 0.003 & 0.026 $\pm$ 0.008 & 0.20 $\pm$ 0.03 \\
\\
\hline
\end{tabular*}
\end{sidewaystable*}

The maximum activity concentration detected was 4.2 $\pm$ 0.6 mBq/m\textsuperscript{3} of \textsuperscript{131}I, which did not include a correction for the volatile iodine components. If the volatile components were included, the activity concentration may have been comparable to the limit of 7.8 mBq/m\textsuperscript{3} set by the Environmental Protection Agency (EPA) although it was well below the limit of 7.4 Bq/m\textsuperscript{3} set by the Nuclear Regulatory Commission (NRC)~\citep{nrc,epa}. The NRC limit is intended to limit public dosage to less than 1 mSv effective dose.

Local weather had a significant impact on measured activities. In particular, sharp decreases in both \textsuperscript{131}I and \textsuperscript{137}Cs activity, shown in Fig.~\ref{time}, can be correlated with rain events in Table~\ref{samplingdata}. 

The measurement of certain isotopes and their ratios allows us to draw some conclusions about the nature and timescale of the Fukushima accident. The measurement of short-lived isotopes, such as \textsuperscript{131,132}I, \textsuperscript{136}Cs and \textsuperscript{132}Te, indicates that radioactivity was released from recently active fuel rods and not from spent fuel cooled for long term. Isotopes such as \textsuperscript{103}Ru, \textsuperscript{141}Ce, \textsuperscript{239}Np and other fission products that were detected after the Chernobyl accident in both Europe and North America~\citep{Dev86,Per90} were not found in the data. Since almost all of the iodine entering containment from a light water reactor cooling system is in the form of CsI which is very soluble~\citep{Bea91}, this suggests that radioactivity due to the release of contaminated steam was far more abundant than from active fuel particles.

Using data from KURF, the ratio of the number of \textsuperscript{134}Cs to \textsuperscript{137}Cs atoms present in filters from the first 10 d of air sampling could also be determined. If the release of radioactivity was from a nuclear weapon, for example, there would be almost no \textsuperscript{134}Cs produced because the formation of \textsuperscript{134}Cs via neutron capture on \textsuperscript{133}Cs is strongly dependent on the length of criticality.
% The number of atoms was calculated as:
% 
% \begin{equation}
% N = \frac{\tau R}{\epsilon_{det}}
% \end{equation}
% 
% \noindent where $\tau$ is the mean lifetime of the relevant isotope. 
The atomic ratio of \textsuperscript{134}Cs to \textsuperscript{137}Cs was found to be 0.070 $\pm$ 0.001, and is consistent with a nuclear reactor accident and not a nuclear weapons test, as expected.

\begin{figure}[htp]
\centering
\includegraphics[width=.75\textwidth, angle=0]{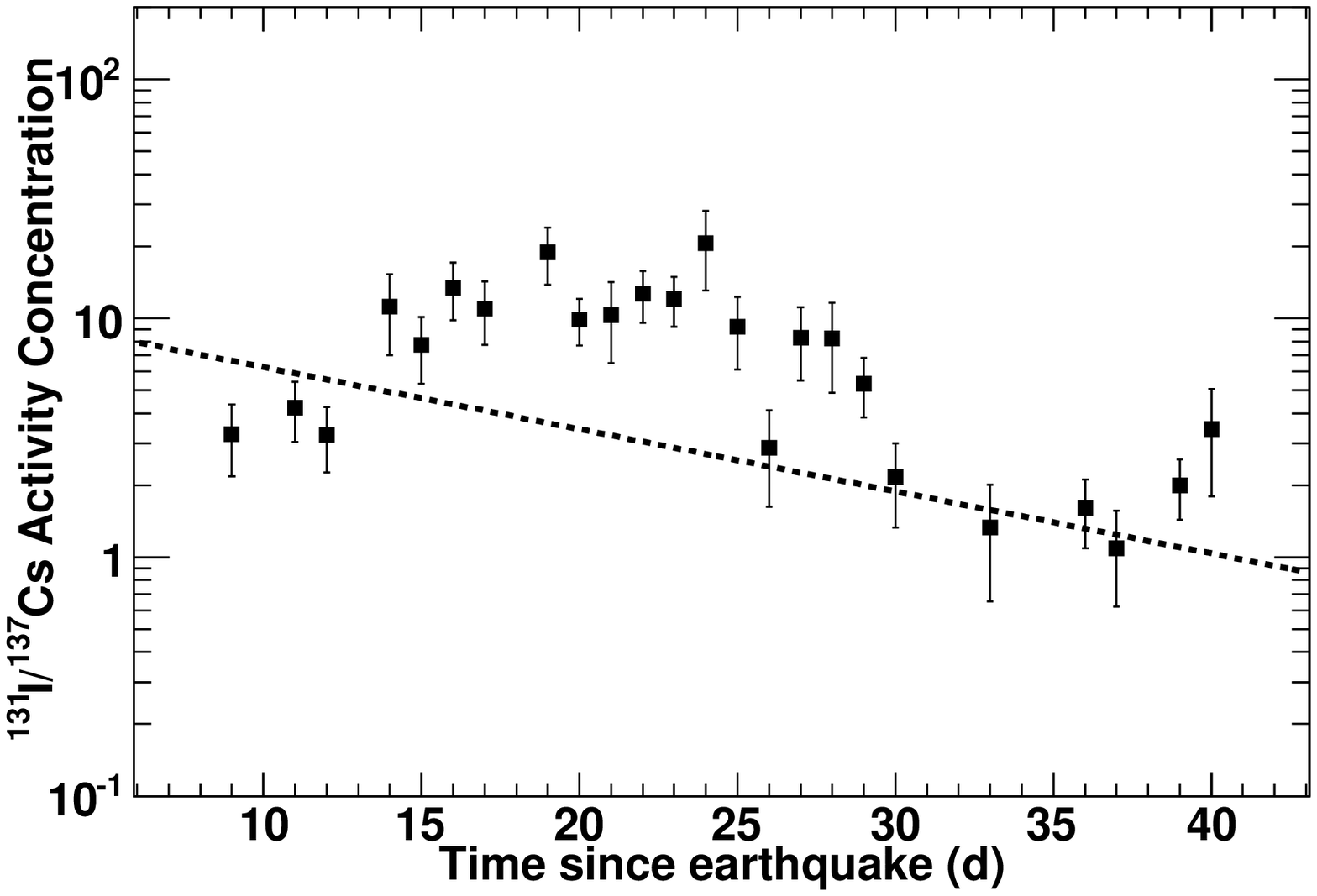}
\caption{The ratio of \textsuperscript{131}I to \textsuperscript{137}Cs activity concentration. The fit is an exponential decay where the decay constant is fixed by the lifetime of \textsuperscript{131}I.}
\label{fig:IdivCs}
\end{figure}

The ratio of \textsuperscript{131}I to \textsuperscript{137}Cs activity from a measurement of seawater samples from the South Discharge Channel, a monitoring post located 330 m south of the Discharge Channel of the unit 1 - 4 reactor sites was investigated by~\citet{Mat11}. The data were described by fitting to an exponential function. The measured ratio of of \textsuperscript{131}I to \textsuperscript{137}Cs activity from data in the current work are shown in Fig.~\ref{fig:IdivCs}. An attempt was made to fit these data to the same exponential function:

\begin{equation}
\frac{\textrm{I}_{activity}}{\textrm{Cs}_{activity}}(t) = \frac{f_\textrm{\scriptsize{I}}}{f_\textrm{\scriptsize{Cs}}}\frac{\tau_\textrm{\scriptsize{Cs}}}{\Delta t \cdot ln(2)}\left(\frac{1}{2}\right)^{t/\tau_\textrm{\scriptsize{I}}}
\end{equation}

\noindent where $f_{\textrm{\scriptsize{I}}}$ and $f_\textrm{\scriptsize{Cs}}$ are the fractions of \textsuperscript{131}I and \textsuperscript{137}Cs per fission ($2.89\times10^{-2}$ and $6.19\times10^{-2}$ respectively~\citep{ipr}), $\tau_\textrm{\scriptsize{I}}$ and $\tau_\textrm{\scriptsize{Cs}}$ are the lifetimes of \textsuperscript{131}I and \textsuperscript{137}Cs and $\Delta t$ is the length of time the reactor has been active. We assume the lifetime of \textsuperscript{131}I is short compared to the reactor operation time, the decay constant is fixed and $\Delta t$ is the only free parameter. The best fit value for $\Delta t$ was found to be 2.6 $\pm$ 0.2 y which is not in good agreement with $\Delta t$ = 1 y reported by~\citet{Mat11}. The data show large deviations from the fit ($\chi^2$/DOF = 4.4 , $p-\textrm{value}$ = $10^{-11}$) which may indicate a difference in the particulate transport mechanisms and efficiencies of the iodine and cesium isotopes.

\section{Conclusions}
Airborne fission products released from the Fukushima Dai-ichi reactor have been measured in Chapel Hill, NC, USA. Fallout from the reactor accident is not expected to have any health implications for the people living in North Carolina or in the United States. The maximum measured \textsuperscript{131}I activity concentration is below limits set by the EPA and NRC as well as many sources of natural background radiation, including radon. After April 20, 2011, the radioactivity in the air fell below our detection limits. The measurements reported here are part of the larger global effort to quantify the transport of fallout from the Fukushima accident. We hope they will contribute to future models of the atmospheric transport of fission products.
 
\section{Acknowledgments}
This work was primarily supported by NSF grant \# PHY 0705014 and DOE grant numbers DE-FG02-97ER4104 and DE-FG02-97ER41033. This research was supported in part by an award from the Department of Energy (DOE) Office of Science Graduate Fellowship Program (DOE SCGF). The DOE SCGF Program was made possible in part by the American Recovery and Reinvestment Act of 2009.  The DOE SCGF program is administered by the Oak Ridge Institute for Science and Education for the DOE. ORISE is managed by Oak Ridge Associated Universities (ORAU) under DOE contract number DE-AC05-06OR23100.  All opinions expressed in this paper are the author's and do not necessarily reflect the policies and views of DOE, ORAU, or ORISE. We acknowledge Lhoist North America for providing us access to the underground site and logistical support. We also thank Mike Miller and Andreas Knecht for useful discussions during the preparation of this manuscript.

\clearpage

%\bibliography{filterbib}

\end{document}